\documentclass{PoS}
\usepackage{graphicx}
\usepackage{amsmath}

\title{Spin Polarizabilities on the Lattice}

\ShortTitle{Spin Polarizabilities on the Lattice}

\author{\speaker{Frank X. Lee}\\
        Physics Department, The George Washington University, Washington, DC, USA\\
        E-mail: \email{fxlee@gwu.edu}}

\author{Andrei Alexandru\\
        Physics Department, The George Washington University, Washington, DC, USA\\
        E-mail: \email{aalexan@gwu.edu}}

\abstract{Spin polarizabilities provide information on the internal structure of hadrons in the presence of weak external electromagnetic fields, and are actively studied by Compton scattering experiments.
They provide finer detail than the regular polarizabilities since they require space and time-varying fields.
Using an effective action in the weak field limit, we have identified methods to isolate 
each of the physical quantities 
($\mu, \alpha, \beta, \gamma_{E1}, \gamma_{M1}, \gamma_{E2}, \gamma_{M2}$) for spin-1/2 hadrons, 
both neutral and charged.
We also perform a lattice QCD simulation to investigate the feasibility of the effective action 
approach.}

\FullConference{XXIX International Symposium on Lattice Field Theory \\
                July 10 - 16 2011\\
                Squaw Valley, Lake Tahoe, California}

\begin{document}

 \section{Introduction}
The interaction of a spin-1/2 hadron with a weak external electromagnetic field can be described 
by an effective, non-relativistic quantum mechanical (QM) action 
in Euclidean space as~\cite{Detmold06} 	
\begin{equation}
S_{QM}=\int d^4x \; L_{QM},
\label{seff}
\end{equation}
where the effective QM Lagrangian is given by
\begin{eqnarray}
L_{QM} &=& \psi^\dagger(x,t)\left[ \left( \partial \over \partial t \right) 
+ (-i\vec{\nabla}-q\vec{A})^2  - \mu \vec{\sigma}\cdot\vec{B} 
+ {1\over 2}\alpha \vec{E}^2 - {1\over 2}\beta \vec{B}^2 
\right. \nonumber \\ & & \left.
+ {i\over 2} \gamma_{E1} \vec{\sigma}\cdot \vec{E}\times \dot{\vec{E}}
- {i\over 2} \gamma_{M1} \vec{\sigma}\cdot \vec{B}\times \dot{\vec{B}}
- {i\over 2} \gamma_{E2} \sigma_i E_{ij} B_j
- {i\over 2} \gamma_{M2} \sigma_i B_{ij} E_j
\right] \psi(x,t).
\label{leff}
\end{eqnarray}
Here 
\begin{equation}
\vec{E}=-{\partial \vec{A} \over \partial t} - \vec{\nabla} A_4, \;\;
\vec{B}=\vec{\nabla}\times \vec{A}, \;\;
\dot{\vec{E}}={\partial \vec{E} \over \partial t}, \;\;
E_{ij}={1\over 2} (\nabla_i E_j + \nabla_j E_i), \hspace{2mm} {\mbox{etc}.}
\label{def}
\end{equation}
and the various physical quantities are 
\begin{eqnarray}
\mu & & \mbox{magnetic moment} 
\nonumber \\
\alpha & & \mbox{electric dipole polarizability} 
\nonumber \\
\beta & & \mbox{magnetic dipole polarizability} 
\nonumber \\
\gamma_{E1} & & \mbox{electric dipole to electric dipole (E1$\rightarrow$E1) spin polarizability} 
          \\
\gamma_{M1} & & \mbox{magnetic dipole to magnetic dipole (M1$\rightarrow$M1) spin polarizability} 
\nonumber \\
\gamma_{E2} & & \mbox{magnetic dipole to electric quadrupole (M1$\rightarrow$E2) spin polarizability} 
\nonumber \\
\gamma_{M2} & & \mbox{electric dipole to magnetic quadrupole (E1$\rightarrow$M2) spin polarizability} 
\nonumber
\label{polar}
\end{eqnarray}
This effective description of the interaction is expected to be valid for fairly weak fields.
The polarizabilities encode rich information on the internal structure of the hadron 
with varying detail. The same polarizabilities appear in low-energy Compton scattering  
amplitudes~\cite{Babusci98}, and are actively pursued by experiments and phenomenological studies. 
Although $\mu$, $\alpha$ and $\beta$ have been studied in lattice QCD~\cite{Tiburzi11}, 
work on spin polarizabilities is just starting~\cite{Engelhardt11}. 
Our goal is to study the feasibility 
of isolating them on the lattice, combining effective theory and lattice QCD. 

 \section{Lattice discretization}
The two-point correlation function in the effective theory (or effective hadron propagator) 
\begin{equation}
G_{ss'}(t,\vec{p},A_\mu)=\int d^3x e^{i\vec{p}\cdot\vec{x}} 
{\int D\psi^\dagger D\psi \psi_s(\vec{x},t) \psi_{s'}^\dagger (0,0) e^{-S_{QM}} \over
 \int D\psi^\dagger D\psi                                    e^{-S_{QM}} }.
\label{energy}
\end{equation}
Since the effective action is in the bilinear form $S_{QM}=\psi^\dagger K \psi$, 
the path integration is easily performed and the resulting correlator $G_{ss'}$ 
is the inverse of the matrix $K$. After projection to finite momentum, 
the correlator is simply a 2 by 2 complex matrix in spin space, 
with $s=1,2$ denoting spin up and down.

We discretize the effective action on a finite lattice of extent $N_x N_y N_z N_t$ and spacing a, 
and evaluate the inverse numerically using a double-precision BiCGSTAB solver 
(since $K$ is not hermitian) with 
a convergence criterion of $10^{-15}$. Since the system is non-relativistic, 
only forward propagation is considered. 
The derivatives are replaced by appropriate differences on the lattice. 
We use the unit system in which 
$\hbar=c=1$ and $e^2=1/137$, and measure all other quantities in terms of fm. 
So the coordinates $(x,y,z,t)$ are in fm, the nucleon mass $m=938/197=4.76$ fm$^{-1}$, 
the nuclear magneton $\mu_N=e/(2m)$ in fm, $E$ and $B$ fields in fm$^{-2}$, 
$\alpha$ and $\beta$ in fm$^3$, and $\gamma$ 's in fm$^4$, etc.

   Most of the results below are obtained on a $24^3\times 64$ lattice at $a=0.1$ fm. 
The source location is at $(12,12,12,4)$. We use Dirichlet boundary condition in the time direction, 
and study different boundary conditions in the spatial directions. 
This is one advantage of the effective approach: we can study boundary conditions exactly 
as they are in lattice QCD. Another advantage is that we can dial individual terms 
in the effective action and see their effects in the correlator. 
No such freedom is afforded in lattice QCD where all relevant terms are present simultaneously. 
Our basic strategy to determine the polarizabilities is: given a lattice QCD correlator, 
an effective QM correlator is computed by adjusting the polarizabilities until a match is found. 

Below we test the effective QM action with some input values 
in Table~\ref{test} (see~\cite{Babusci98}). 
We want to see if the input values could be recovered as output from the effective correlator alone. 
Such testing is useful because the same methodology can be used on lattice QCD correlators.
The advantage here is that the effective QM correlator is fast to compute and free from Monte Carlo noise.
Once a method is identified to isolate them, we can determine their true values 
by matching the effective QM correlator with the corresponding lattice QCD one.
\begin{table}[htb] 
\caption{Proton and neutron values for testing purposes.}
\label{test}
\begin{tabular}{ccccccccc}
\hline\hline
 & m & $\mu$ & $\alpha$ & $\beta$ & $\gamma_{E1}$ & $\gamma_{M1}$ & $\gamma_{E2}$ & $\gamma_{M2}$ \\
 & (fm$^{-1}$) & $(\mu_N)$ & $(10^{-4} \mbox{fm}^3)$ & $(10^{-4} \mbox{fm}^3)$ & $(10^{-4} \mbox{fm}^4)$& $(10^{-4} \mbox{fm}^4)$& $(10^{-4} \mbox{fm}^4)$ & $(10^{-4} \mbox{fm}^4)$ \\
\hline
$p$  & 4.76 & 2.79 & 12 & 2 & -3.4 & 2.7 & 1.9 & 0.3  \\
\hline
$n$  & 4.76 &-1.91 & 12 & 2 & -5.6 & 3.8 & 2.9 &-0.7  \\
\hline\hline
\end{tabular}
\end{table}

 \section{Free field}
In the case of periodic spatial boundary conditions, which we denote as bc=(1,1,1,0) 
(the boundary condition in time is always Dirichlet in this study), 
the correlator projected to zero momentum is a single exponential $G(t)=w \exp(-\lambda t)$ 
where the spectral weight is $w=1/(1+a m)/a^3$ and the mass of the particle 
is related to $\lambda$ by $\lambda=\ln(1+a m)$. This was confirmed numerically.
In the case of Dirichlet in $x$ and periodic in $y$ and $z$, 
denoted as bc=(0,1,1,0), the correlator contains a tower of states with discrete energies
\begin{equation}
G(t)=\sum_{n=1}^{N_x} w_n e^{-\lambda_n t},\;\;\mbox{where}\;\; \lambda_n=\ln(1+E_n) 
\;\;\mbox{and}\;\; E_n=m+{p_n^2\over 2m}.
\label{gfree}
\end{equation}
Here $p_n$ is the discrete lattice momentum given by
\begin{equation}
p_n={\sin(n\pi a/(L_x+a)) \over a/2} \;\;\mbox{with}\;\; n=1,3,5,\cdots,
\label{gfree2}
\end{equation}
where $L_x$ is the lattice size in x direction.
Note that the even terms are absent under the specific boundary conditions.
Since both $w_n$ and $\lambda_n$ can be calculated exactly in the free-field case, 
we are able to confirm numerically that the correlator is indeed described by 
Eq.~(\ref{gfree}) and Eq.~(\ref{gfree2}) to double precision. 
Similar checks were performed for other boundary conditions used in this work:
bc=(0,0,1,0), bc=(0,1,0,0), and bc=(1,0,0,0).

 \section{Constant electric field: $\alpha$}
Next we turn on the electric field.
The vector potential $\vec{A}=(-iEt,0,0)$ in Euclidean space (we set $A_4=0$) 
corresponds to a real-valued electric field E in the x direction in Minkowski space. 
We impose Dirichlet boundary condition in the x direction to eliminate the unphysical 
wrapping-around-the-lattice effects~\cite{Alexandru10} present in lattice QCD so our boundary 
condition in this case is bc=(0,1,1,0). For neutron, the interaction part of the effective action 
(all terms except the first two in Eq.~(\ref{leff})) is $L_{int}=-1/2\alpha E^2$. 
We checked that $\alpha$ can be recovered straightforwardly from the correlator for different fields. 
For proton, the electric field causes an acceleration term in the correlator, 
in addition to the $\alpha$ term. We checked that $\alpha$ can be cleanly extracted from the 
unpolarized ratio 
$G(t,\alpha)/G(t,\alpha=0)$, which not only cancels out the acceleration term, 
but also the common energy term $m+p^2/(2m)$. The advantage here is that we do not need 
to know the details of the acceleration: its effects are fully incorporated in the effective 
QM correlator. Since the acceleration is the same in the effective theory and QCD, 
its effect can be canceled out in the same way in QCD in the ratio 
$G_{QCD}(t,\alpha)/G_{QM}(t,\alpha=0)$. This is an alternative method to the one used in previous 
studies of charged hardons~\cite{Tiburzi09}.

 \section{Constant magnetic field: $\mu$ and $\beta$}
Now we turn on the magnetic field.
The vector potential $\vec{A}=(0,Bx,0)$ corresponds to a real-valued magnetic field B 
in the z direction in Minkowski space. The quantization condition $ea^2B=n(2\pi/N_x)$ 
would ensure constant B on the periodic lattice, but the fields produced are too strong: 
the lowest quantized B=306 fm$^{-2}$ on our lattice would cause a proton energy shift of 
$1.5$ GeV from $\mu B$ (out of $0.938$ GeV). One could reduce the strength from another quantization 
condition  $ea^2B=n(2\pi/N_xN_y)$ by `patching' up the field on the edge of the 
lattice~\cite{Rubinstein95} to ensure 
uniform flux through every plaquette, but the lowest field (12.8 fm$^{-2}$) is still too strong 
for our purposes. So we abandon quantization and use weak fields (as small as B=0.2 fm$^{-2}$). 
We impose Dirichlet bc in x and put the source in the center of lattice to minimize the effects 
from breaking the quantization condition, and Dirichlet bc in y to eliminate the unphysical 
wrapping-around-the-lattice effects, so bc=(0,0,1,0). For neutron, the interaction in the effective 
action is $L_{int}=-\mu B\sigma_z-1/2\beta B^2$. We checked that $\mu$ can be recovered 
from the ratio of polarized correlators $G_{11}(t)/G_{22}(t)$ and $\beta$ can be recovered 
from the product $G_{11}(t)*G_{22}(t)$. For proton, there is the additional complication of 
Landau levels. We checked that the extraction of $\mu$ is unaffected since the Landau-level 
effects cancel out in the ratio. 
For $\beta$, the Landau-level effects can be cleanly removed by taking the double ratio 
\begin{equation}
R_{\beta}=[G_{11}(t,\beta)G_{22}(t,\beta)]/[G_{11}(t,\beta=0)G_{22}(t,\beta=0)].
\label{Rbeta}
\end{equation}

 \section{Time and space varying fields: $\gamma$ 's}
For $\gamma_{E1}$, the choice of the vector potential
\begin{equation}
\vec{A}=(e_1t^2/a, ie_2t, 0),
\label{Age1}
\end{equation}
where $e_1$ and $e_2$ are real parameters,
gives a time-varying electric field 
\begin{equation}
\vec{E}=(-2e_1t/a, -ie2, 0).
\label{Ege1}
\end{equation}
Because the vector potential has components in both x and y directions, Dirichlet boundary conditions 
are imposed in these directions to eliminate any wrapping-around-the-lattice effects. 
The boundary conditions are denoted as bc=(0,0,1,0).
The interaction Lagrangian has the form
\begin{equation}
L_{int}= 1/2\alpha (4t^2e_1^2/a^2-e_2^2)+\gamma_{E1}e_1e_2\sigma_z/a. 
\label{Lge1}
\end{equation}
Due to the $\sigma_z$ dependence $\gamma_{E1}$ can be isolated from the ratio of polarized 
correlators 
\begin{equation}
R_{\gamma_{E1}}=G_{11}/G_{22}\sim \exp(2\gamma_{E1}e_1e_2t/a), 
\label{Rge1}
\end{equation}
despite the complicated $\alpha$ term. 
For proton, the electric field causes an additional acceleration term, but the ratio is not 
affected and $\gamma_{E1}$ can be extracted the same way. 
In both cases (neutron and proton), the method works out nicely numerically. 
We also checked analytic continuation: $\vec{A}=(e_1t^2/a, ie_2t, 0)$ and $\exp(2\gamma_{E1}e_1e_2t/a)$ 
 give identical results for $\gamma_{E1}$ as $\vec{A}=(e_1t^2/a, e_2t, 0)$ and 
$\exp(2i\gamma_{E1}e_1e_2t/a)$ for small $e_1$ and $e_2$ values.

For $\gamma_{E2}$, the choices are
\begin{equation}
\vec{A}=(e_1y/2, 0, ie_2t z/a), \;\;\\
\vec{E}=(0,0, -ize_2/a), \;\;\\
\vec{B}=(0,0, -e_1/2).
\label{AEBge2}
\end{equation}
Dirichlet boundary conditions are imposed in the x and z directions in this case so bc=(0,1,0,0).
The interaction Lagrangian has the form
\begin{equation}
L_{int}= 1/2\mu e_1\sigma_z -1/2\alpha z^2e_2^2/a^2-1/8\beta e_1^2+1/4\gamma_{E2}e_1e_2\sigma_z/a.
\label{Lge2}
\end{equation}
 The unpolarized $\alpha$ and $\beta$ terms can be eliminated in the ratio $G_{11}/G_{22}$ 
which is left with both $\mu$ and $\gamma_{E2}$ terms. Fortunately, the $\mu$ term is linear 
in the fields and $\gamma_{E2}$ term quadratic. So the $\mu$ contribution can be eliminated 
by averaging the mass shifts over $(e_1,e_2)$ and $(-e_1,-e_2)$. 
At the correlator level, the eliminations can be achieved simultaneously by the double ratio 
\begin{equation}
R_{\gamma_{E2}}=[G_{11}(e_1,e_2)G_{11}(-e_1,-e_2)]/[G_{22}(e_1,e_2)G_{22}(-e_1,-e_2)]
\sim \exp(1/2\gamma_{E2}e_1e_2t/a).
\label{Rge2}
\end{equation}
The method applies to both charged and uncharged hadrons.
It also implies that in order to determine $\gamma_{E2}$ in QCD 
we need to perform two lattice QCD calculations: one with an original set of fields, 
the other with the fields reversed. 

For $\gamma_{M2}$, the field choices are 
\begin{equation}
\vec{A}=(0, e_1t^2/a, ie_2t/2), \;\;\\
\vec{E}=(0, 0, -ie_2/2), \;\;\\
\vec{B}=(-e_1x/a, 0, e_1z/a)
\label{AEBgm2}
\end{equation}
Dirichlet boundary conditions are imposed in the y and z directions in this case so bc=(1,0,0,0).
The interaction Lagrangian 
\begin{equation}
L_{int}=\mu e_1 (x\sigma_x-z\sigma_z)/a-1/8\alpha e_2^2-1/2\beta e_1^2 (x^2+z^2)/a^2 -1/4\gamma_{M2}e_1e_2\sigma_z/a 
\label{Lgm2}
\end{equation}
has a similar form as that for $\gamma_{E2}$, so the same methodology can be used to extract $\gamma_{M2}$. 
Namely, the double-double ratio in Eq.~(\ref{Rge2}) 
would be proportional to $\exp(-1/2\gamma_{M2}e_1e_2t/a)$. 

Finally, for $\gamma_{M1}$, no choice of the vector potential can isolate it independently. 
We found that the choices 
\begin{equation}
\vec{A}=(0, ie_1tz/a, e_2x), \;\;\\
\vec{E}=(0, -ie_1z/a, 0), \;\;\\
\vec{B}=(-ie_1t/a, -e_2, 0),
\label{AEBgm1}
\end{equation}
result in an interaction 
\begin{equation}
L_{int}=\mu (ie_1t\sigma_x/a+e_2\sigma_y) -1/2\alpha z^2e_1^2-1/2\beta (-e_1^2t^2/a^2 +e_2^2) -1/4(2\gamma_{M1}-\gamma_{E2})e_1e_2\sigma_z/a,
\label{Lgm1}
\end{equation}
that has contributions from both $\gamma_{M1}$ and $\gamma_{E2}$. 
We can isolate the combination ($2\gamma_{M1}-\gamma_{E2}$) in the same way as $\gamma_{M2}$, 
then use the previously determined $\gamma_{E2}$ on the same lattice to pin down $\gamma_{M1}$. 
Dirichlet boundary conditions are imposed in the y and z directions in this case so bc=(1,0,0,0).

In summary, these numerical tests demonstrate that all four spin polarizabilities 
can be disentangled by a judicious choice of the vector potential, 
and manipulation of effective correlators. 
The real challenge is whether the same can be done with lattice QCD correlators which are born with Monte Carlo noise.

 \section{$\gamma_{E1}$: a lattice QCD case study}	
\begin{figure}[tbh]
   \centering
   \includegraphics[width=6.0in]{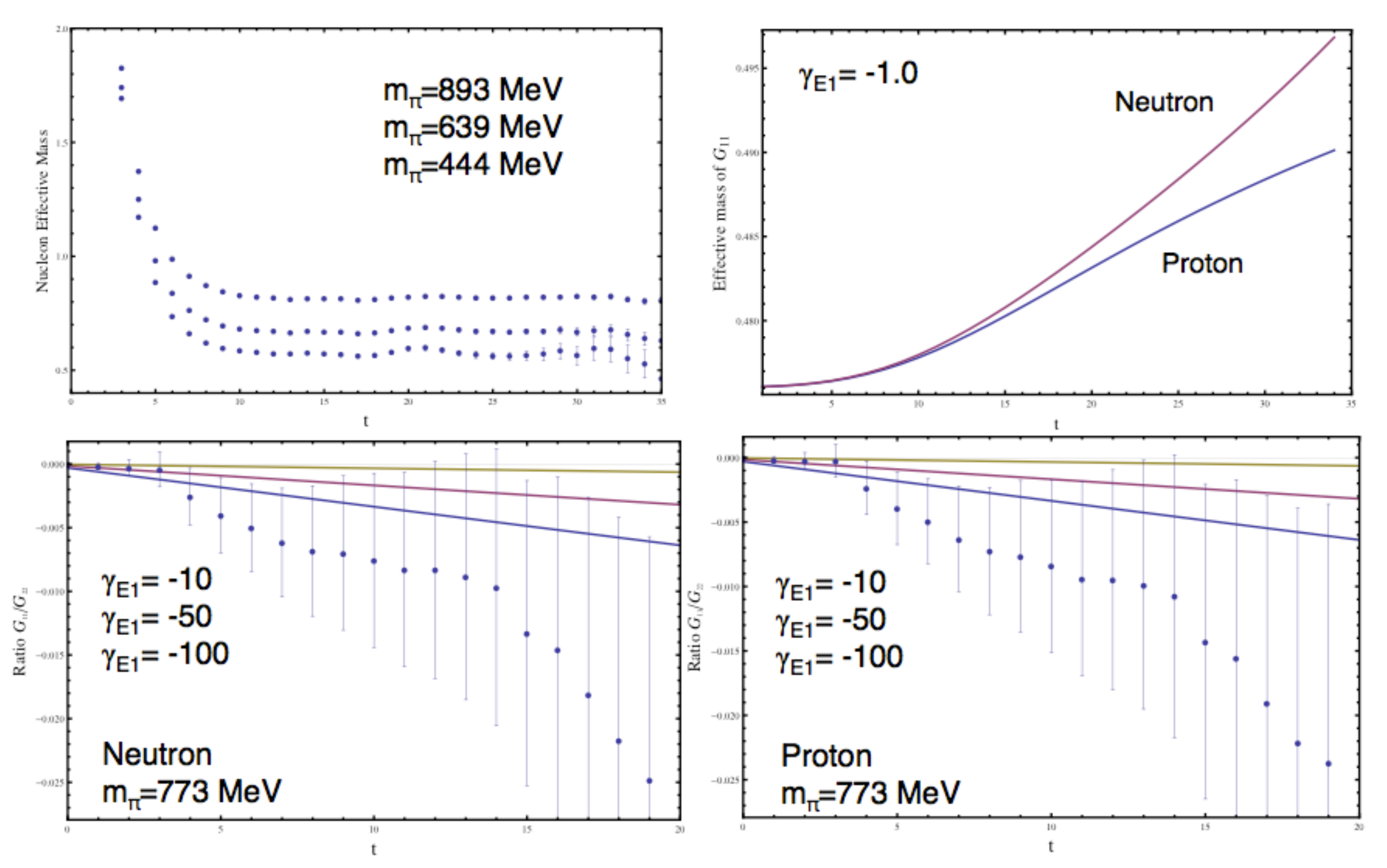} 
   \caption{Some results from the $\gamma_{E1}$ study.}
   \label{ge1}
\end{figure}

Here we perform a preliminary real lattice QCD simulation to see whether the strategy proposed 
is feasible. The QCD data are generated on $24^3\times 48$ lattice with Wilson actions at $a$=0.093 fm. 
The field is applied by $\vec{A}=(e_1t^2/a, e_2t, 0)$ and bc=(0,0,1,0), 
with field values $(e_1,e_2) = (0.23, 0.46)$ fm$^{-2}$. 
We analyzed 700 configurations for 6 pion masses from 893 to 404 MeV. In Figure~\ref{ge1}, 
we see that the nucleon ground state starts to dominate at $t=7$. 
The effective mass of $\mbox{Re}[G_{11}(t)]$ is very different in neutron and proton 
(mostly due to acceleration effects of the proton), but the ratio 
$\mbox{Im}[G_{11}(t)/G_{22}(t)] \sim \sin[\gamma_{E1}e_1e_2t/a]$ is the same. 
The same ratio in QCD suggests a negative $\gamma_{E1}$ for both neutron and proton, 
but suffers from large statistical uncertainties. The 3 lines are predictions at 3 different 
values of $\gamma_{E1}$ in units of $10^{-4}$ fm$^4$. 
The QCD data suggest a wide range for $\gamma_{E1}$ (anywhere between 0 and -200). 
To get a more definitive value for $\gamma_{E1}$,  the error bars need to be reduced by 
at least a factor 10, which means the statistics need to be improved by 100 times if no 
other improvements are made.

 \section{Conclusion}
The effective QM correlator provides a controlled and fast way of examining spin polarizabilities 
and their systematics on the lattice. By designing the background fields and manipulating 
the correlators, we have identified methods to isolate all the physical quantities
($\mu, \alpha, \beta, \gamma_{E1}, \gamma_{M1}, \gamma_{E2}, \gamma_{M2}$) for spin-1/2 hadrons 
as defined in Eq.~(\ref{leff}).
This is possible because of the different field and spin 
dependences in the terms containing these quantities. 
By matching with the corresponding lattice QCD correlators 
on the same lattice with the same boundary conditions,  
their QCD values can be determined in principle. 
A preliminary study of $\gamma_{E1}$ using lattice QCD data 
suggests that the challenge lies in the Monte Carlo noise present
in the lattice QCD simulations: the noise must be reduced significantly before reliable information 
can be extracted.
Several strategies are being explored to this end.

\section*{Acknowledgment}
This work is supported in part by U.S. Department of Energy
under grant DE-FG02-95ER40907. 
The computing resources at NERSC, JLab, and GW IMPACT have been used.

\end{document}